# The Lens-Thirring effect in the anomalistic period of celestial bodies

Ioannis Haranas

*Dept. of Physics and Astronomy, York University, 4700 Keele Street, Toronto, Ontario, M3J IP3, Canada*
e-mail: ioannis@yorku.ca

Omiros Ragos

*Dept. of Mathematics, University of Patras, GR-26500 Patras, Greece*
e-mail: ragos@math.upatras.gr

Ioannis Gkigkitzis

*Dept. of Mathematics, East Carolina University, 124 Austin Building, East Fifth Street Greenville, NC 27858-4353, USA*
e-mail: gkigkitzisi@ecu.edu

**Abstract** In the weak field and slow motion approximation, the general relativistic field equations are linearized, resembling those of the electromagnetic theory. In a way analogous to that of a moving charge generating a magnetic field, a mass–energy current can produce a gravitomagnetic field. In this contribution, the motion of a secondary celestial body is studied under the influence of the gravitomagnetic force generated by a spherical primary. More specifically, two equations are derived to approximate the periastron time rate of change and its total variation over one revolution (i.e., the difference between the anomalistic period and the Keplerian period). Kinematically, this influence results to an apsidal motion. The aforementioned quantities are numerically estimated for Mercury, the companion star of the pulsar PSR 1913+16, the companion planet of the star HD 80606 and the artificial Earth satellite GRACE–A. The case of the artificial Earth satellite GRACE–A is also considered, but the results present a low degree of reliability from a practical standpoint.

## 1 Introduction

In the weak-field linearization of the general relativistic field equations, the theory predicts that the gravitational field of a rotating primary body results to a magnetic-type force that is called gravitomagnetic. This force affects secondary bodies, gyroscopes and clocks that move around this primary as well as electromagnetic waves [1]. Lens and Thirring [2] worked on the gravitomagnetic influence on the motion of test particles orbiting a slow rotating primary mass.

They particularly interested on celestial systems. They found that this influence, hereinafter called Lens–Thirring effect, results to precession in the argument of the perigee and the ascending node of the orbits of the secondaries.

Far from the rotating body, the gravitomagnetic field can be expressed as [3]:

$$\vec{B}_{gm}(r) = \frac{G}{cr^3}\left[3(\hat{r}\cdot\vec{S})\hat{r} - \vec{S}\right] \quad (1)$$

where $G$ is the constant of universal gravitation, $c$ is the speed of light, $\hat{r}$ is the unit vector along the position vector $\vec{r}$ and $\vec{S}$ is the angular momentum vector of the rotating body. The field indicated by Eq. (1) equals to that of a magnetic dipole whose moment is [4]:

$$\vec{\mu}_{gm} = \frac{G\vec{S}}{c} \quad (2)$$

Then, a secondary, that is moving with a velocity $\vec{v}$ in this field, is affected by a non-central acceleration that is given by:

$$\vec{A} = -2\left(\frac{\vec{v}}{c}\right) \times \vec{B}_{gm} \quad (3)$$

In stronger gravitational fields, like those of neutron stars and black holes, higher order corrections in $\vec{v}/c$ become important for the orbits of test particles, too. These kinds of scenarios have been studied by Cappoziello et al. [5] and, also, Schäffer [1].

The time rate of change, which is caused by the Lens–Thirring effect, in the argument of the periastron and the ascending node of the secondary are given by the following relations [6]:

$$\frac{d\omega}{dt} = -\frac{6GS\cos i}{c^2 a^3 (1-e^2)^{3/2}} \quad (4)$$

$$\frac{d\Omega}{dt} = \frac{2GS}{c^2 a^3 (1-e^2)^{3/2}} \quad (5)$$

where $a$, $e$, $i$ represent the semimajor axis, the eccentricity and the inclination of the orbit. The radial, transverse and normal to the orbital plane components of the acceleration given in Eq. (3) are [4]:

$$A_R = \xi \cos i (1 + e\cos f) \quad (6)$$

$$A_T = -\xi e \sin i \sin f \qquad (7)$$

$$A_N = \xi \sin i \left[ 2\sin f \left(1 + e\cos f\right) + e\sin f \cos f \right] \qquad (8)$$

where $f$ is the true anomaly, and $\xi$ is equal to:

$$\xi = \frac{2GS}{c^2} \sqrt{\frac{G\mathcal{M}(1-e^2)}{a}} \frac{(1+e\cos f)^3}{a^3 (1-e^2)^4}. \qquad (9)$$

Here $\mathcal{M}$ denotes the mass of the primary.

In this paper, we deal with the Lens–Thirring effect on the relative orbit of the secondary body with respect to the primary one. In other words, we reduce the initial two-body problem to a central-force problem. Within this framework, We establish the periastron time rate of change induced by this effect. Then, we determine the perturbation in the anomalistic period (the time interval elapsed between two successive transits of the secondary at the periastron) via the difference between this one and that predicted by the Kepler's theory [7],[8]. Numerical applications of our formulae are given for the motions of Mercury, the companion star of the pulsar PSR 1913+16, the companion planet of the star HD 80606 as well as the artificial Earth satellite GRACE–A.

## 2 Rate of change and variation per revolution of the periastron time

Consider the unperturbed relative orbit of the secondary, obviously a Keplerian ellipse. Let $n$ be its mean motion, and $M$ the mean anomaly. First, we will express the rate of change of the periastron time $T_0$ in terms of the true anomaly $f$. The well-known relation

$$M = n(t - T_0). \qquad (10)$$

connects $M$ to $T_0$. Here $t$ denotes the time variable. We differentiate Eq. (3) with respect to $t$ to obtain:

$$\frac{dT_0}{dt} = 1 + \frac{(t-T_0)}{n}\frac{dn}{dt} - \frac{1}{n}\frac{dM}{dt}.. \qquad (11)$$

The rate of change of the mean motion $dn/dt$ can be found by using Kepler's third law. On the unperturbed Keplerian ellipse this law is expressed as follows:

$$n^2 a^3 = GM. \tag{12}$$

Therefore:

$$\frac{dn}{dt} = -\frac{3n}{2a}\frac{da}{dt}. \tag{13}$$

Substituting Eq. (13) in Eq. (11) we obtain that:

$$\frac{dT_0}{dt} = 1 + \frac{3(t-T_0)}{2a}\frac{da}{dt} - \frac{1}{n}\frac{dM}{dt}. \tag{14}$$

In the presence of a perturbation, the rates of change of the orbital elements can be found by using Gauss' planetary equations. In our case and for the semimajor axis, the mean anomaly, the argument of the periastron and the ascending node, they read [9]:

$$\frac{da}{dt} = \frac{2}{n\sqrt{1-e^2}} \left[ e \sin f \, A_R + \frac{a(1-e^2)}{r} A_T \right]. \tag{15}$$

$$\frac{dM}{dt} = n - \frac{2}{na}\left(\frac{r}{a}\right) A_R - \sqrt{1-e^2}\left(\frac{d\omega}{dt} + \cos i \frac{d\Omega}{dt}\right). \tag{16}$$

$$\frac{d\omega}{dt} = \frac{\sqrt{1-e^2}}{nae}\left[-A_R \cos f + A_T \left(1 + \frac{r}{a(1-e^2)}\right)\sin f\right] - \cos i \frac{d\Omega}{dt}. \tag{17}$$

$$\frac{d\Omega}{dt} = \frac{1}{na\sqrt{1-e^2}\sin i} A_N \left(\frac{r}{a}\right) \sin(\omega + f). \tag{18}$$

Substituting Eqs. (15)-(18) into Eq. (14) and, after doing some algebra, Eq. (14) simplifies to:

$$\frac{dT_0}{dt} = \left[\frac{2r}{n^2 a^2} - \frac{(1-e^2)}{n^2 ae}\cos f + \frac{3e(t-T_0)}{na\sqrt{1-e^2}}\sin f\right] A_R$$
$$+ \left[\frac{(1-e^2)}{n^2 ae}\left(1 + \frac{r}{a(1-e^2)}\right)\sin f + \frac{3(t-T_0)\sqrt{1-e^2}}{nr}\right] A_T \tag{19}$$

Next, we express Eq. (19) in terms of the eccentric anomaly $E$ by using the well-known relations (see e.g., [10]):

$$r = a(1 - e\cos E), \tag{20}$$

$$\frac{dE}{dt} = \frac{n}{1 - e\cos E}, \tag{21}$$

$$t - T_0 = \frac{E - e\sin E}{n}, \tag{22}$$

$$\cos f = \frac{\cos E - e}{1 - e\cos E} = \frac{a(\cos E - e)}{r}, \tag{23}$$

$$\sin f = \frac{\sqrt{1-e^2}\sin E}{1-e\cos E} = \frac{a\sqrt{1-e^2}\sin E}{r}. \tag{24}$$

By using Eqs (20) and (24), we can rewrite $A_R$ and $A_T$ in the following way

$$A_R = \frac{2GS}{c^2}\sqrt{\frac{GM(1-e^2)}{a}}\frac{\cos i}{a^3(1-e\cos E)^4}, \tag{25}$$

$$A_T = \frac{2GS}{c^2}\sqrt{\frac{GM}{a}}\frac{e\sin i \sin E}{a^3(1-e\cos E)^4}. \tag{26}$$

Then, Eq. (19) takes the form:

$$\frac{dT_0}{dE} = \frac{1 - e\cos E}{a^3 n}\left[A_R B(e,E) - A_T C(e,E)\right], \tag{27}$$

where

$$B(e,E) = 2(1-e\cos E) - \frac{(1-e^2)(\cos E - e) + 3e(E - e\sin E)}{1 - e\cos E}, \tag{28}$$

and

$$C(e,E) = \frac{\sin E\sqrt{1-e^2}}{e(1-e\cos E)}\left[(2 - e^2 - e\cos E) + 2e(E - e\sin E)\right], \tag{29}$$

To proceed with the integration we expand in power series of the eccentricity the following terms that appear in Eq. (27):

$$\frac{1}{(1-e\cos E)^4} \cong 1 + 4e\cos E + 10e^2\cos^2 E + 20e^3\cos^3 E + 35e^4\cos^4 E + O(e^5), \tag{30}$$

$$\frac{1}{(1-e\cos E)^5} \cong 1 + 5e\cos E + 15e^2\cos^2 E + 35e^3\cos^3 E + 70e^4\cos^4 E + O(e^5). \tag{31}$$

If we substitute Eqs. (30)-(31) into (27), take into account that $S = \frac{2}{5}MR^2 w$ and integrate over one period, we obtain:

$$\Delta T_0 = \frac{\pi \mathcal{R}^2 w}{40 n^3 c^2} \left( \frac{G^3 \mathcal{M}^3 (1-e^2)}{a^9} \right)^{1/2} \begin{bmatrix} (128\cos i + 64\sin i) \\ +e(64\pi \sin i - 256\cos i) \\ +96 e^2 (\sin i - \cos i) \\ +32 e^3 ((5\pi - 2)\sin i - 23\cos i) \\ +120 e^4 (\sin i - 5\cos i) + O(e^5) \end{bmatrix}, \quad (32)$$

where $\mathcal{R}$ and $w$ denote the radius and the angular velocity of the primary, respectively. Ii is seen that $\Delta T_0$ is proportional to $\mathcal{M}^{3/2} \mathcal{R}^2 w$. This is how the anomalistic period change of the secondary depends on the characteristics of the primary.

In particular, for polar orbits ($i = 90^\circ$), Eq. (32) becomes:

$$\Delta T_{0p} = \frac{\pi \mathcal{R}^2 w}{10 n^3 c^2} \left( \frac{G^3 \mathcal{M}^3 (1-e^2)}{a^9} \right)^{1/2} \left( 16 + 16\pi e + 24 e^2 + 8(5\pi - 2) e^3 + 30 e^4 \right) \quad (33)$$

while, for equatorial ones ($i = 0^\circ$), it takes the form:

$$\Delta T_{0e} = -\frac{\pi \mathcal{R}^2 w}{10 n^3 c^2} \left( \frac{G^3 \mathcal{M}^3 (1-e^2)}{a^9} \right)^{1/2} \left( 32 - 64 e - 24 e^2 - 184 e^3 - 150 e^4 \right), \quad (34)$$

In the special case of polar circular orbits we obtain that:

$$\Delta T_{0p} = \frac{16 \pi R^2 w}{10 n^3 c^2} \left( \frac{G^3 M^3}{a^9} \right)^{1/2}, \quad (35)$$

and for equatorial circular orbits we get:

$$\Delta T_{0e} = -\frac{32 \pi \mathcal{R}^2 w}{10 n^3 c^2} \left( \frac{G^3 \mathcal{M}^3}{a^9} \right)^{1/2}. \quad (36)$$

For elliptical orbits, Eqs. (33) and (34) produce the following relation between equatorial and polar anomalistic times:

$$\Delta T_{0e} = -\frac{32 - 64 e - 24 e^2 - 184 e^3 - 150 e^4}{16 + 16\pi e + 24 e^2 + 8(5\pi - 2) e^3 + 30 e^4} \Delta T_{0p}. \quad (37)$$

while, for circular ones, Eqs. (35) and (36) give that:

$$\Delta T_{0e} = -2 \Delta T_{0p}. \quad (38)$$

## 3 Numerical results

First, we proceed with the calculation of the change of the perihelion time of Mercury. We have used for the orbital parameters of this planet the following values: $a = 57909083$ km, $e = 0.205$, $i = 7.004^\text{o}$, $n = 8.26 \times 10^{-7}$ rad/s. For Sun we have used that. $\mathcal{M} = 1.99 \times 10^{30}$ kg, $w = 2.863 \times 10^{-6}$ rad/s, while, supposing that this star is a spherical body, its radius is, approximately, $\mathcal{R} = 6.96 \times 10^8$ km. Then, Eq. (32) predicts the following anomalistic period change for Mercury due to the Lens–Thirring effect:

$$\Delta T_0 = 9.824 \times 10^{-5} \text{ s/rev.} \tag{39}$$

Figure 1 represents the variation of the anomalistic period time of change for Mercury by altering the values of its semimajor axis and eccentricity. Figure 2 depicts the corresponding variation if its inclination is considered to be equal to 0.

Next, we consider the case of the pulsar PSR 1913+16 and its companion star. For the pulsar we have used that $\mathcal{M} = 2.870 \times 10^{30}$ kg, $w = 0.454881162$ rad/s and $\mathcal{R} = 9.74 \times 10^3$ km, while for the orbital parameters the companion $a = 1.950 \times 10^6$ km, $e = 0.617$, $i = 45^\text{o}$ and $n = 1.60608 \times 10^{-4}$ rad/s. Then, the change of the anomalistic period of this companion is found to be:

$$\Delta T_0 = 6.073 \times 10^{-8} \text{ s/rev.} \tag{40}$$

Similarly, assuming that the companion of PSR 1913+16 has an orbital inclination $i = 0^\text{o}$, we obtain:

$$\Delta T_{0\text{e}} = -3.112 \times 10^{-7} \text{ s/rev.} \tag{41}$$

while, for $i = 90^\text{o}$, we have that:

$$\Delta T_{0\text{p}} = 3.971 \times 10^{-7} \text{ s/rev.} \tag{42}$$

Figure 3 represents the variation of the anomalistic period time of change for Mercury by altering the values of its eccentricity and inclination.

Another test case will be the well known extra-solar planet $b$, a companion planet of the star HD 80606 in the constellation of Ursa Major. This planet is a superjovian planet whose orbital parameters (http://exoplanet.hanno-rein.de/system.php?id=HD+80606+b) are approximately: $a = 6.795 \times 10^7$ km, $e = 0.933$, $i = 89.285^\text{o}$ and $n = 4.0 \times 10^{-8}$ rad/s. Also, the mass, the angular velocity and the radius of HD 80606 are $\mathcal{M} = 1.791 \times 10^{30}$ kg, $w = 1.168 \times 10^{-3}$ rad/s and $\mathcal{R} = 5.99 \times 10^{-8}$ km. Then, the change of the anomalistic period of the planet $b$ will be:

$$\Delta T_0 = 3.910 \times 10^{-6} \text{ s/rev.} \qquad (43)$$

Assuming $i = 0^\text{o}$ for the inclination of $b$, we get that:

$$\Delta T_{0e} = -3.518 \times 10^{-6} \text{ s/rev.} \qquad (44)$$

while, for $i = 90^\text{o}$, Eq. (32) predicts that:

$$\Delta T_{0p} = 3.956 \times 10^{-6} \text{ s/rev.} \qquad (45)$$

Figure 4 represents how this period change varies, assuming different values for its semimajor axis and eccentricity. Figure 5 depicts the corresponding variation if its inclination was equal to 0.

Comparing the values of $\Delta T_0$ for the polar and equilateral orbit in any of the two last example cases, it seems that this change for polar orbits is, in absolute value, bigger than that for equilateral ones.

In the case of GRACE–A, an artificial satellite of Earth, we have that its orbital elements are $a = 6876.4816$ km, $e = 0.00040989$, $i = 89.025446^\text{o}$ and $n = 0.001100118$ rad/s (http://www.csr.utexas.edu/grace/). For Earth we have that $\mathcal{M} = 5.9736 \times 10^{24}$ kg and $w = 7.292 \times 10^{-5}$ rad/s and $\mathcal{R} = 6378.1363$ km. Then, for GRACE–A, the change of the anomalistic period is:

$$\Delta T_0 = 1.867 \times 10^{-7} \text{ s/rev.} \qquad (46)$$

But we must say that, practically, these results have a low degree of reliability, because of the very small eccentricity of this satellite. It is known that for quasicircular orbits the position of the periastron (hence the periastron passage time) cannot be accurately determined. However our calculations are of some interest regarding the magnitude order of the perigee passage time variation.

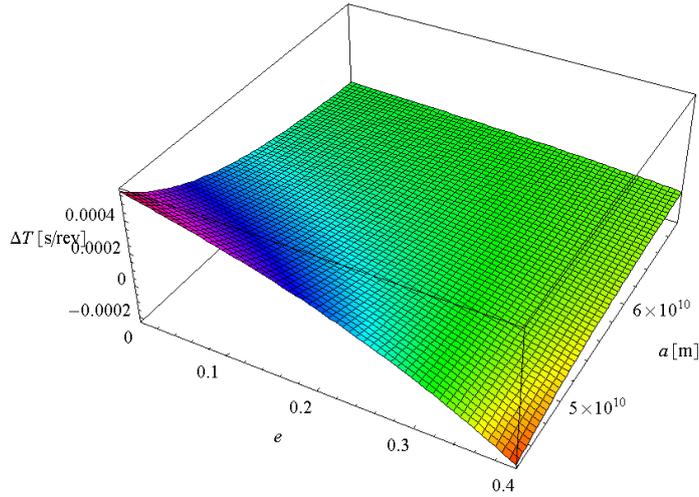

**Fig.1.** The anomalistic period time change of Mercury due to Lens–Thirring effect as a function of *e* and *a* by supposing that their values are altered in the ranges [0,1] and [perihelion, aphelion], respectively.

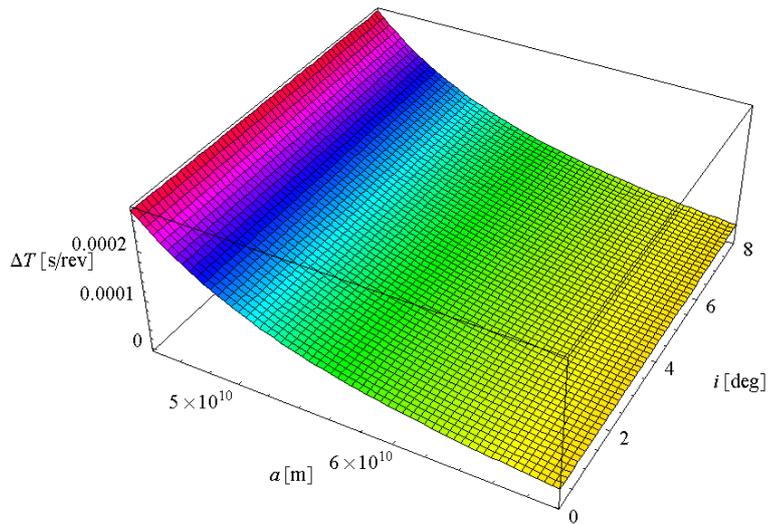

**Fig.2.** The anomalistic period time change of Mercury due to Lens–Thirring effect as a function of *a* and *i* by supposing that their values are altered in the ranges [perihelion-aphelion] and [0,8], respectively.

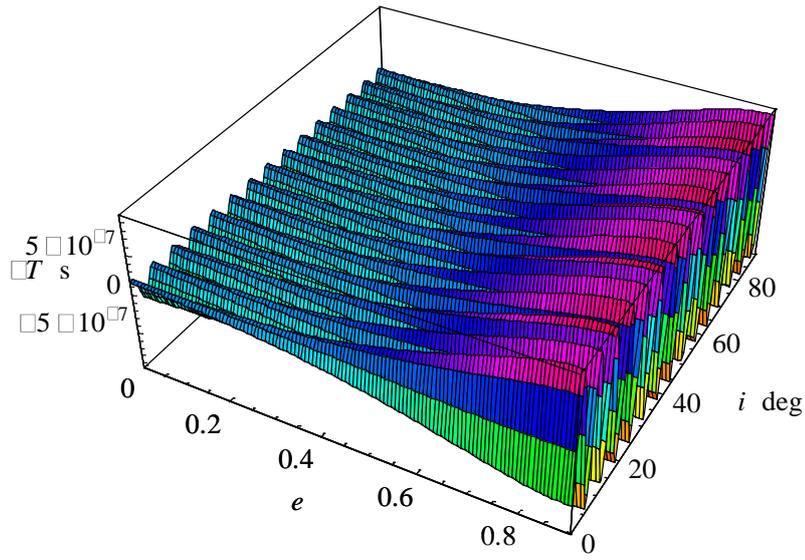

**Fig.3.** The anomalistic period time change of the companion star of the pulsar 1913+16 due to Lens–Thirring effect as a function of *e* and *i* by supposing that their values are altered in the ranges [0,1] and [0,90], respectively.

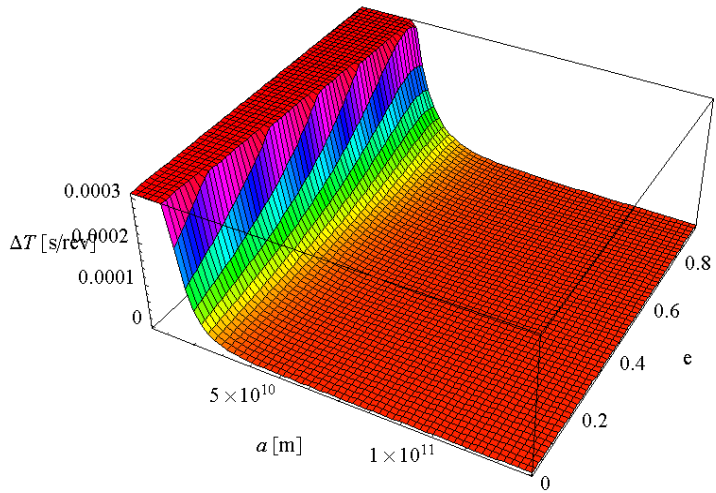

**Fig.4.** The anomalistic period time change of the planet HD 80606 *b* due to Lens–Thirring effect as a function of *e* and *a* by supposing that their values are altered in the ranges [0.1] and [periastron-apoastron], respectively.

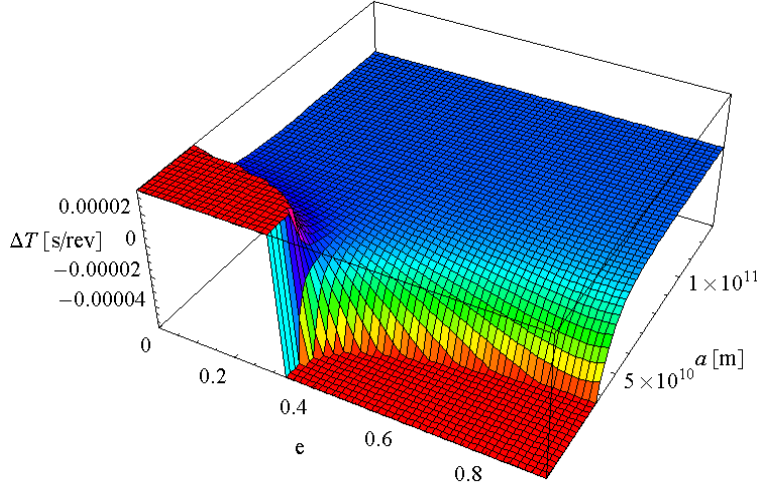

**Fig.5.** The anomalistic period time change of the planet HD 80606 *b* due to Lens–Thirring effect as a function of *e* and *a* by supposing that their values are altered in the ranges [0,1] and [periastron-apoastron], respectively. An equatorial orbit case (*i* = 0) is considered.

## 4 Summary and concluding remarks

We use the components of the gravitomagnetic force produced by a primary celestial body that gives rise to the Lens–Thirring effect in order to derive an eccentric anomaly–dependent equation that estimates the rate of change of the periastron time $T_0$ of a secondary body orbiting this primary. By using the integral of this equation over a whole revolution, we have found that the Lens–Thirring effect contributes to an advance or a recess of the periastron time, depending on the inclination and eccentricity of the secondary. A variation for $T_0$ of the order of microseconds may be detectable by today's technology. This variation was estimated for some concrete astronomical cases. Mercury exhibits an advance of 98.24 μs, the PSR 1913+16 companion star an advance of 0.061 μs, the planet HD 80606 *b* an advance of 3.91 μs, and finally in the case of GRACE–A this advancing effect is equal to 0.1867 μs. The presented results can constitute a possible test for the action of the Lens–Thirring force on the solar system bodies, or other celestial objects. Of course, this is not the only effect to be considered. For example, other effects like general relativistic and quantum effects can be also considered. This is another topic for us to deal with in the nearest future.